\documentclass[useAMS,usenatbib]{mn2e}
\usepackage{graphicx}
\usepackage{txfonts}
\usepackage{natbib}

\title[QPOs and energy spectra from the two brightest ULXs in M~82]{Quasi-Periodic Oscillations and energy spectra from the two brightest Ultra-Luminous X-ray sources in M~82}

\author[Caballero-Garc\'{i}a, Belloni \& Zampieri]{M. D. Caballero-Garc\'{i}a$^{1}$\thanks{E-mail:
mcaballe@brera.inaf.it}, T. Belloni$^{1}$, L. Zampieri$^{2}$\\
$^{1}$ INAF-Osservatorio Astronomico di Brera, Via E. Bianchi 46, I-23807 Merate (LC), Italy \\
$^{2}$ INAF-Osservatorio Astronomico di Padova, Vicolo dell'Osservatorio 5, 35122 Padova, Italy 
}

\begin{document}


\pagerange{\pageref{firstpage}--\pageref{lastpage}} \pubyear{2002}

\maketitle

\label{firstpage}

\begin{abstract}
Ultra-Luminous X-ray sources are thought to be accreting black holes that might host
Intermediate Mass Black Holes (IMBH), proposed to exist by
theoretical studies, even though a firm detection (as a class) is still missing. The brightest 
ULX in M~82 (M~82 X--1) is probably one of the best candidates to host an IMBH. In this work we analyzed the data
of the recent release of observations obtained from M~82 X--1 
taken by {\it XMM-Newton}. We performed a study of the timing and spectral properties of
the source. We report on the detection of ${\approx}(46{\pm}2)$\,mHz Quasi-Periodic Oscillations (QPOs) in the power 
density spectra of two observations. A comparison
of the frequency of these high-frequency QPOs with previous detections 
supports the 1:2:3 frequency distribution as suggested in other studies. We discuss the implications 
if the ${\approx}(46{\pm}2)$\,mHz QPO detected in M~82 X--1 is the fundamental
harmonic, in analogy with the High-Frequency QPOs observed in black hole binaries.
For one of the observations  we have detected for the first time a QPO at 8\,mHz (albeit at a
low significance), that coincides with a hardening of the spectrum. We suggest that the QPO
is a milli-hertz QPO originating from the close-by transient ULX M~82 X--2, with analogies to the 
Low-Frequency QPOs observed in black hole binaries.

\end{abstract}


\section{Introduction}

Ultra-Luminous X-ray sources (ULXs) are point-like, off-nuclear, extra-galactic sources, with observed X-ray luminosities
(${\rm L}_{\rm X}{\ge}10^{39}\,{\rm erg}\,{\rm s}^{-1}$) higher than the Eddington luminosity for a stellar-mass black-hole
(${\rm L}_{\rm X}{\approx}10^{38}\,{\rm erg}\,{\rm s}^{-1}$). The true nature of these objects is still debated
\citep{feng11,fender12}. Although the definition of ULXs encompasses different types of sources,
the majority of them are likely to be accreting BHs in binary systems.
However, there is still no unambiguous estimate for the mass of the compact
object hosted in these systems. Assuming an isotropic emission, in order to avoid the violation of the Eddington limit, ULXs
might be powered by accretion onto Intermediate Mass Black Holes (IMBHs) with masses in the range
$10^{2}-10^{4}\,{\rm M}_{\odot}$ \citep{colbert99}. It has been also suggested that ULXs appear very luminous
because of beaming \citep{king01}, super-Eddington emission \citep{begelman02}, a combination of these effects \citep{poutanen07}
or because they contain a moderately high mass BH (\citealt{zampieri09,mapelli09}).

An approach to study the nature of ULXs is through time variability. The analysis of the aperiodic variability in the X-ray
flux of X-ray binaries is a powerful tool to study the properties of the inner regions of the accretion disc around compact objects
(for a review see \citealt{vdk05}). In particular, Quasi-Periodic Oscillations (QPOs) provide well-defined frequencies, which can be
linked to specific time scales in the disc. QPOs in Black Hole Binaries (BHBs) can be broadly divided into three classes: (i) QPOs at very low frequencies 
($<0.02$\,Hz), probably associated to oscillations and instabilities in the accretion disc (see \citealt{mo97,b97,b00}); (ii) Low-Frequency 
QPOs (LFQPOs), with typical frequencies between 0.1 and 10\,Hz, probably connected to similar oscillations in neutron star systems (see
e.g. \citealt{b02,r02a,vdk05,casella05}), over whose origin there is no consensus; in Black Hole Binaries (BHBs) 3 main different types of
LFQPOs have been identified (\citealt[][and references therein]{casella05}); (iii) High-Frequency QPOs (HFQPOs), with a typical
frequency of ${\approx}35-450$\,Hz, in a few cases observed to appear in pairs \citep{st01a,st01b,belloni12,belloni13b}. It is currently unclear whether these QPOs 
have constant frequencies and whether they have special ratios \citep{belloni13a}. However, since they identify the highest frequencies observed in these systems, they are the best candidates
for association with, e.g., the Keplerian frequency at the innermost stable orbit.
Whatever their physical nature, as they originate in the inner regions of accretion discs around black holes, these features are expected to
be produced also in ULXs. However, if ULXs contain IMBHs of $(10^2-10^4)\,{\rm M}_\odot$, the frequencies are correspondingly smaller.

In this paper we report on the results from a joint spectral and timing analysis of five $20-30$\,ks {\it XMM-Newton} observations of
M~82 performed on March-November 2011. The plan of the paper is the following. In Sec.~\ref{observ},~\ref{analysis} we present the timing and spectral analysis of
the data. Sec.~\ref{results} reports the detection of QPOs in the data, together with a comparison with previous findings. The results from 
the spectral analysis and the relationship with the timing properties are also shown. Finally, in Sec.~\ref{discuss} we discuss our results. 

\subsection{M~82 X--1}

M~82 X--1 (also named CXOU~J095550.2+694047) is one of the brightest ULXs in the sky. Its host galaxy is the prototype of starburst galaxy and is located nearby, at a distance
of 3.9\,Mpc \citep{sakai99}, making M~82 X--1, with an average X-ray luminosity ${\gtrsim}10^{40}$\,${\rm erg\,s^{-1}}$, an excellent
target for performing X-ray studies. The position of M~82 X--1 is within $1\arcsec$ of the position of the infrared source and super star cluster MGG~11
\citep{kaaret04,portegies04}. Simulations of the dynamical evolution of the cluster MGG~11 show that stellar collisions in its extremely
dense core may have led to the formation of an IMBH \citep{portegies04}. \citet{strohmayer03} found QPOs
in the range 50--110~mHz which they identified as arising from M~82 X--1 (and later confirmed by \citealt{feng07} using {\it Chandra} data). If the QPO frequency 
is associated with the orbital frequency around a non-rotating black hole, then it would imply a mass limit of $2{\times}10^{4}\,{\rm M}_{\odot}$. The QPOs 
were discovered in {\it XMM-Newton} data and confirmed in {\it RXTE} data \citep{strohmayer03}. The QPOs are only occasionally detected in the
{\it RXTE} data and no apparent correlation between QPO detections and the source flux level has been found. Based on a comparison of the spectral and
timing properties of stellar-mass black hole X-ray binaries, \citet{fiorito04} estimated a mass of the order of $10^3{\rm M}_{\odot}$ for
the compact object producing the QPOs. The method uses the photon index-QPO frequency correlation that has been seen in BHBs \citep{vignarca03,titarchuk04}. The 
QPOs were confirmed in a longer {\it XMM-Newton} observation, where their frequency was 113~mHz \citep{mucciarelli06,dewangan06}. Also \citet{casella08} estimated
the mass of the BH to be in the range $(100-10^3){\rm M}_{\odot}$. Their method relies in the location of the ULX in the variability plane of BHs, that relates
the mass of the BH with its timing properties. 

\subsection{M~82 X--2}

The X-ray source M~82 X--2 (also named CXOU~J095551.4+694043) was identified as a transient ULX in M~82 from multiple X-ray observations with {\it Chandra} \citep{kaaret06,feng07,kong07}. 
It lies on the sky plane at $5''$ from M~82 X--1, which is most of the time the brightest source in M82. Only {\it Chandra} is able to resolve them in X-rays. From 
{\it Chandra} observations it has been seen that M~82 X--2 is sometimes the second brightest X-ray source in M~82, but it is undetected during the rest of the time \citep{matsumoto01,feng07}.
\citet{feng07} argued that M~82 X--2 is more likely to be an IMBH than a stellar-mass object accreting from a massive star according to its transient nature and high outburst-luminosity 
(${\approx}10^{40}$\,${\rm erg\,s^{-1}}$; \citealt{kalogera04}). \citet{kaaret06} reported significant
timing noise near 1\,mHz from M~82 X--2 in one {\it Chandra} observation that later \citet{feng10} confirmed it in the form of a milli-hertz QPO. Identifying the frequency of this QPO
to that of a LFQPO and applying a linear mass-frequency scaling relationship \citet{feng10} derived a value of ${\approx}10^4{\rm M}_{\odot}$ for the mass of the BH.

\section{Observations and data reduction} \label{observ}

In this work we analyzed the recently publicly available observations of M~82 X--1 collected by the {\it XMM-Newton} satellite (see Tab.~\ref{log_obs} for details). 
The EPIC camera was operating in the {\it Full Frame} mode and with the {\it Medium Filter} set. 

For the timing analysis, we filtered the EPIC pn+MOS event files, selecting only the best-calibrated events (pattern${\le}4,12$ for the pn and
the MOS, respectively), and rejecting flagged events (i.e. keeping only flag$=0$ events) from a circular region on the source (centre at coordinates ${\rm RA}=09\,h55\,m50.2\,s$,
${\rm Dec}=+69\,d40\,m47\,s$; \citealt{mucciarelli06}) and radius $13\,$arcsec. Many gaps in the light curves were present. These gaps are of instrumental origin, due 
to periods of intense background flaring. How we dealt in the analysis with these time periods is shown in Sec.~\ref{timing}. No background subtraction was applied, to preserve 
the statistical properties of the distribution of powers in the power spectrum. We 
payed particular attention to extract the list of photons not randomized in time. To this purpose, we used the tasks {\it epchain} for the pn and
{\it emproc} (with {\it randomizetime=no}) for the MOS cameras, respectively.

For the spectral analysis we used only the EPIC pn camera, in order to avoid issues due to cross-calibration effects. Additionally, the EPIC pn camera
has a higher effective area (i.e. double) than each one of the MOS cameras and has sufficient statistics for the fit. We applied the standard filtering 
of removing time periods with high count-rates in the FOV in the 10-12\,keV energy range (EPIC-pn only). We filtered the event files, 
selecting only the best-calibrated events (pattern${\le}4$ for the pn), and rejecting flagged events (flag$=0$). We extracted 
the flux from a circular region on the source centred at the coordinates of the source and radius $13\,$arcsec (the same as the region used for the timing analysis). The background was extracted from
a circular region (with a radius of $18\,$arcsec), centred in a position in the same chip, far away from the boundaries and not far from the source.
We built response functions with the {\it Science Analysis System} (SAS) tasks {\tt rmfgen} and {\tt arfgen}. We
fitted the background-subtracted spectra with standard spectral models
in XSPEC 12.7.0 \citep{a1}. All errors quoted in this work are $68\%$ ($1{\sigma}$) confidence. The spectral fits were limited to the 0.3-10\,keV range, where the
calibration of the instruments is the best. The spectra were rebinned in order to have at least 25 counts for each background-subtracted spectral channel and to avoid oversampling of the
intrinsic energy resolution by a factor larger than 3 \footnote{As recommended in: \\
http://xmm.esac.esa.int/sas/current/documentation/\\
threads/PN\_spectrum\_thread.shtml}.

\begin{table}
 \centering
 \begin{minipage}{120mm}
  \caption{Log of the observations from this work.}
  \label{log_obs}
  \begin{tabular}{@{}cccc@{}}
  \hline
   Obs. Number               &   Obs. Date  &    MJD    &  Exposure Time~$^1$  (seconds)             \\
 \hline
   1  &   2011--03--18   &   55\,638    &     26\,657      \\
   2  &   2011--04--09   &   55\,660    &     23\,836      \\
   3  &   2011--04--29   &   55\,680    &     28\,219      \\
   4  &   2011--09--24   &   55\,828    &     22\,843      \\
   5  &   2011--11--21   &   55\,886    &     23\,914      \\
\hline
\end{tabular}
\footnotetext{$^1$ The on-time of an observation. }
\end{minipage}
\end{table}

\section{Analysis}  \label{analysis}

\subsection{Timing Analysis}  \label{timing}

We performed an analysis of the fast time variability of M~82 X--1 in the 1-10\,keV energy range. 
The (0.3-1\,keV) and (1-10\,keV) EPIC non-background subtracted count rates are ${\approx}0.20\,{\rm cts}\,{\rm s}^{-1}$ and  
$(1-2)\,{\rm cts}\,{\rm s}^{-1}$, respectively. Since the (0.3-1\,keV) count rate is very low we ignored this energy range in the 
analysis and limited it to the (1-10\,keV) energy range since it has better statistics. The time resolution of the instruments 
is 2.6\,s (MOS) and 73.4\,ms (pn). 

The pn+MOS light curve was binned at the lowest time resolution of the two (2.6\,s). This yields a Nyquist frequency of $=0.19$\,Hz.
We produced Power Density Spectra (PDS) using intervals of 512 bins in each light curve. The PDS were then averaged 
together for each observation. Many gaps (153 and 56 with a duration of ${\le}100$\,s each, which represent ${\approx}50,25\%$ of the total time of the light curves, respectively) were present in the 
light curves of Obs.~2,~3 and were filled with a Poissonian realization around the local mean value of counts before and after the gap. The PDS were normalized according to 
\citet{leahy83}. The resulting total PDS was logarithmically rebinned with the bin increase with frequency adjusted on a case by case basis, in order to improve the statistics. All 
of the PDS show low-frequency flat-topped noise.

PDS fitting was carried out with the standard XSPEC fitting package by using a unit response. Fitting the (1-10\,keV) PDS with a model composed 
by a Lorentzian for the low-frequency noise and a constant for the Poisson noise results in
non acceptable chi-square values (${\chi}^{2}/{\nu}=87/61,143/100,170/100,57/41$) for Obs.~1-4. The width and the fractional rms 
of the zero-centred Lorentzian component are $(3-70)$\,mHz and $(10-18)\%$ for Obs.~1-4, respectively. In the case of Obs.~5
this model is a good description of the data (${\chi}^{2}/{\nu}=101/100$). The width and the fractional rms of the zero-centred 
Lorentzian component are $1.3_{-0.8}^{+2.0}$\,mHz and $(11.8{\pm}1.4)\%$ 
in this observation. The PDS of Obs.~3 shows positive 
residuals at low-frequencies that we fitted with an additional power-law component, yielding an improvement of the fit
of ${\Delta}{\chi}^{2}=20$ for ${\nu}=2$.

\begin{figure*}
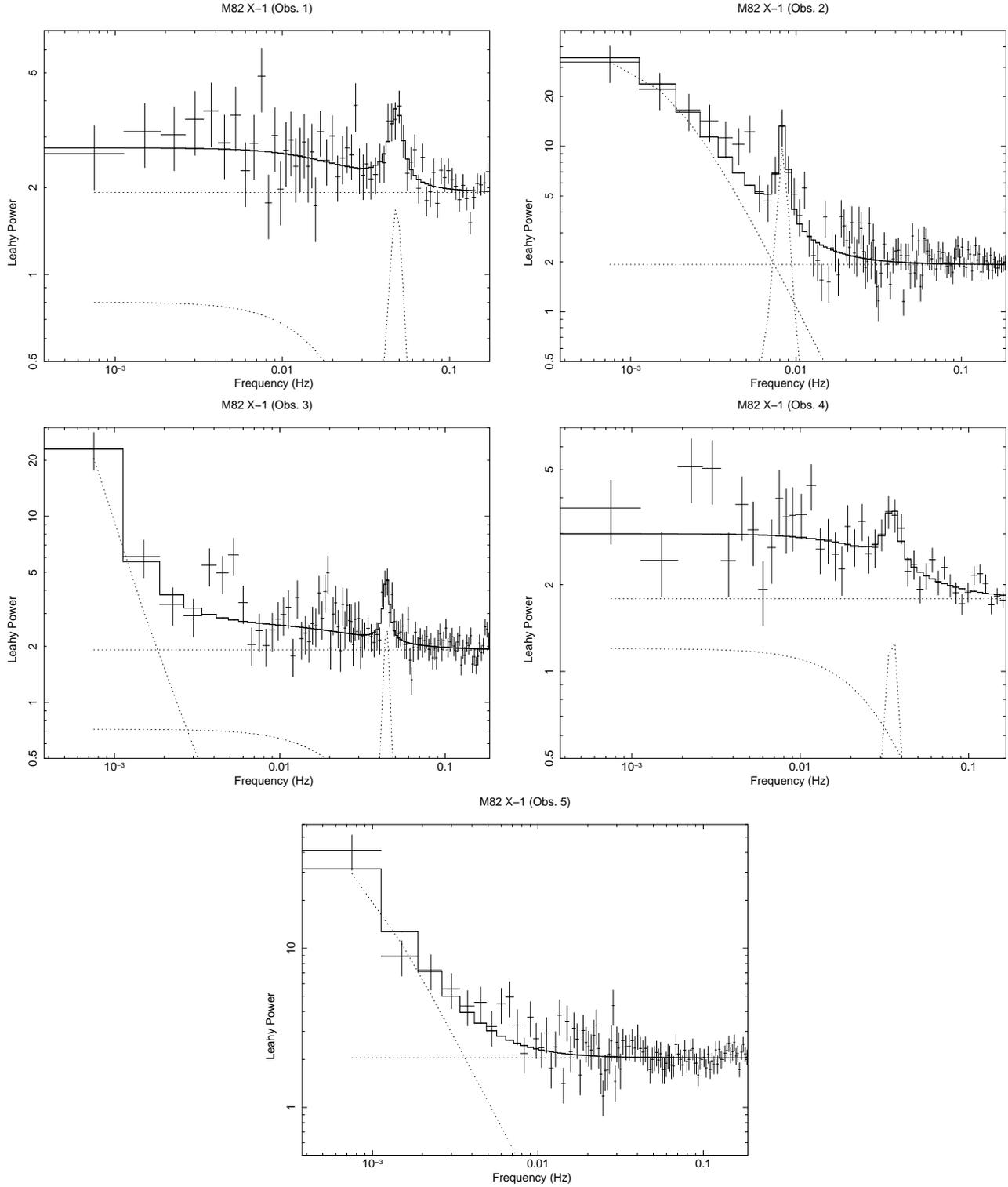

\centering
 \includegraphics[bb=34 5 570 700,width=6.7cm,angle=270,clip]{pds1.ps}
 \includegraphics[bb=34 5 570 700,width=6.7cm,angle=270,clip]{pds2.ps}
 \includegraphics[bb=34 5 570 700,width=6.7cm,angle=270,clip]{pds3.ps}
 \includegraphics[bb=34 5 570 700,width=6.7cm,angle=270,clip]{pds4.ps}
 \includegraphics[bb=34 5 570 700,width=6.7cm,angle=270,clip]{pds5.ps}
 \caption{Power density spectra from the {\it XMM-Newton} EPIC/pn data used in this work in the (1--10\,keV) energy band and $(4{\times}10^{-4}-0.2)$\,Hz frequency range during observations 1--5 (top-left to bottom-right)
with the best fit model (solid line) and the model components (dotted line).} 
 \label{plots_timing}
\end{figure*}

With the continuum model adopted, the (1-10\,keV) PDS of Obs.~1-4 still show positive residuals 
centred at ${\nu}_{\rm QPO}{\approx}50,8,40,30\,{\rm mHz}$. We added a Lorentzian component to fit these excesses. The detection of this peak is significant (${\ge}3.6{\sigma}$, calculated 
as the value of the Lorentzian normalization divided by its $1{\sigma}$ error) except for Obs.~4 ($<3{\sigma}$). The integrated 1-10\,keV
fractional variability of these QPOs are ${\approx}8-10\%$. The final fits
result in mostly acceptable chi-square values ${\chi}^{2}/{\nu}=61/58,127/97,119/95,43/38,101/100$ for Obs.~1-5, respectively. 
The PDS are plotted in Fig.~\ref{plots_timing} and the parameters from the fits and the QPOs obtained are in Tab.~\ref{table_timing} \footnote{We 
have replaced the data with white noise at the same count rate level and repeated the PDS analysis. We do not detect QPOs at the frequencies
we report in the paper, indicating that they are not caused by the presence of filled data gaps. Additionally, in order to test whether the 
QPOs are caused by background variability, we have extracted light curves from background regions and 
repeated the analysis and have seen that no QPO is detected.}.

\begin{table*}
 \centering
 \begin{minipage}{120mm}
  \caption{Values obtained from the timing analysis.}
  \label{table_timing}
  \begin{tabular}{@{}lccccc@{}}
  \hline
                                &   Obs.~1                      &          Obs.~2        & Obs.~3                  &    Obs.~4             &    Obs.~5              \\
 \hline
   Count rate\,$^{1}$\,(${\rm cts}\,{\rm s}^{-1}$)   & $2.39{\pm}0.02$    & $2.77{\pm}0.02$     & $2.10{\pm}0.02$      & $2.33{\pm}0.02$      & $3.15{\pm}0.02$  \\
   ${\nu}_{\rm lor}$\,(mHz)     &   $0$                     &  $0$                 &   $0$                   &   $0$                 &  $0$                   \\
   FWHM\,(mHz)                  &  $50{\pm}20$              &  $3.3{\pm}0.6$       &  $60{\pm}20$            & $70{\pm}20$           &  $1.3_{-0.8}^{+2.0}$   \\
   Norm.                        &  $0.06{\pm}0.03$          &  $0.21{\pm}0.03$     &  $0.06{\pm}0.02$        & $0.13{\pm}0.04$       &  $0.13_{-0.06}^{+0.14}$ \\
   Fractional rms ($\%$)        &  $10.5{\pm}1.4$           &  $17.8{\pm}1.2$      &  $11.8{\pm}1.6$         & $15.4{\pm}0.9$        &  $11.8{\pm}1.4$                 \\
   ${\nu}_{\rm QPO}$\,(mHz)     &  $48.1{\pm}1.1$           & $8.3{\pm}0.2$        &   $44.1{\pm}0.5$        &  $35.2{\pm}1.2$       &  --                    \\
   ${\rm FWHM}$\,(mHz)          &  $9{\pm}4$                & $0.9{\pm}0.4$        &  $4{\pm}2$              &  $7{\pm}4$            &  --                   \\
   ${\rm Norm.}$                &  $0.026{\pm}0.007$        & $0.016{\pm}0.004$    &  $0.016{\pm}0.004$      &  $0.015{\pm}0.006$    &  --                  \\
   Fractional rms ($\%$)        &  $10.2{\pm}1.4$           & $7.6{\pm}0.9$        &  $9{\pm}2$              &  $8.0{\pm}1.8$        &  --                  \\
   ${\rm Significance}_{\rm QPO}$\,(${\sigma}$)~$^2$ &   $3.6$                      &  $3.8$               &  $4.5$               &   $2.6$               &    --            \\
   ${\Gamma}_{\rm poisson}$     &   $0$                     &  $0$                 &   $0$                   &   $0$                 &  $0$                   \\
   ${\rm Norm.}_{\rm poisson}$  &  $1.93{\pm}0.07$          & $1.93{\pm}0.03$      &  $1.92{\pm}0.06$        &   $1.78{\pm}0.10$     &  $2.05{\pm}0.03$       \\
   ${\Gamma}$                   &  --                       & --                   &  $2.3{\pm}0.4$          &   --                  &  --                    \\
   Norm.                        &  --                       & --                   &  ${\le}10^{-6}$         &   --                  &  --                    \\
   ${\chi}^{2}/{\nu}$           &   $61/58$                 & $127/97$             &  $119/95$               &   $43/38$             &  $101/100$             \\
\hline
\end{tabular}
\footnotetext{Values for the count rate and characteristics of the noise (using the model {\tt lorentz+lorentz+powerlaw+powerlaw}) for the six observations.
The PDS were created in the (1--10)\,keV energy band and $(4{\times}10^{-4}-0.2)$\,Hz frequency range. Errors are $68\%$ confidence errors.}
\footnotetext{$^1$ Background-subtracted count rate from the pn+MOS cameras. }
\footnotetext{$^2$ Single trial significance of the QPO.  }
\end{minipage}
\end{table*}

\subsection{Spectral Analysis}  \label{spectr}

We started fitting the spectra with an absorbed power-law model, using the Tuebingen-Boulder ISM absorption model ({\tt tbabs} in XSPEC) to account
for the interstellar absorption (${\rm N}_{\rm H}=4{\times}10^{20}\,{\rm cm}^{-2}$ in the direction to M~82). This parameter was set free to vary
in order to account for intrinsic absorption. We fitted the spectra simultaneously, constraining the column density to be the same for all the spectra. With
this model we obtained a bad description of the spectra (${\chi}^{2}/{\nu}{\gg}2$ with ${\nu}=781$ d.o.f.), with positive residuals at ${\le}2$\,keV and high-energy curvature at ${\gtrsim}5$\,keV. The spectra are curved with a
break at ${\approx}5-7$\,keV, in agreement with what has been found in previous studies from a sample of ULXs (see e.g. \citealt{stobbart06,gladstone09,caballero10,caballero11}).

To account for the low-energy residuals, that are in the form of excesses at low energies (around 1 and 2\,keV) we had to include an {\tt apec} model, which accounts for the diffuse X-ray emission from the galaxy. We fixed the metal abundances to
solar, in agreement with a previous work based on {\it XMM-Newton} data \citep{read02}. We obtained a better
fit (${\chi}^{2}/{\nu}{\approx}2.0$, with ${\nu}=779$). Some residuals in the form of excesses at around 2\,keV were still present. In a different context, 
\citet{stevens03} tried to fit the 
nuclear emission of M~82 with two components coming from the diffuse emission of the galaxy. Therefore, we included a second {\tt apec} model but this was not enough to improve neither the spectral fit (${\chi}^{2}/{\nu}{\approx}2.0$, 
with ${\nu}=782$) nor the appearance of the residuals. The parameters of the {\tt apec} components were constrained to be the same between the observations. Following \citet{stevens03,mucciarelli06,ranalli08} we tried to fit the 
residual galactic emission with a dual {\tt apec} model with different absorbing columns. That improved 
the fit (${\Delta}{\chi}^{2}{\approx}390$ for 3 d.o.f.) and flattened the low-energy residuals. Nevertheless, the fit was still poor (${\chi}^{2}/{\nu}=1.9$ with ${\nu}=775$). The column densities obtained are in agreement
with those obtained from {\it Chandra} observations at the position of the ULX \citep{fengkaar10}.

To account for the high-energy residuals we added a cut-off, replacing the power-law component by 
a high-energy cut-off ({\tt highecut*powerlaw } in XSPEC) model to fit the high-energy spectra. This model improved the fit substantially (${\chi}^{2}/{\nu}=1.16$, with ${\nu}=765$).
We included a multicolor inner disc emission component ({\tt diskbb} in XSPEC), as previously done by \citet{mucciarelli06}. This improved the fit statistics by ${\Delta}{\chi}^2=43$ for ${\nu}=10$ d.o.f. (i.e. a $4{\sigma}$ improvement). 
We also tried by substituting the cut-off power-law by the more physical {\tt compTT} model. The temperature for the input photons from the Comptonization component was set equal to the inner
disc temperature. We also tried to fit the spectra not fixing the inner disc temperature to the temperature of the input photons from the Comptonization component and 
obtained a non-significant improvement of the fit (i.e. $<3{\sigma}$). Therefore we kept the constraint. With this model we obtained very similar quality of the fits (${\chi}^{2}/{\nu}{\approx}1.15$, with ${\nu}=760$). The 
improvement of the fit by adding the disc component ({\tt diskbb} in XSPEC) 
was found to be significant at the $3.6{\sigma}$ level. The resulting parameters are in agreement with previous studies for a sample of ULXs \citep{gladstone09}.

The most relevant results of this spectral analysis and the derived unabsorbed fluxes are
in Tab.\,\ref{log_spe} and in Fig.~\ref{plot_spe}. The errors on the total flux (plus the flux from every individual component) were calculated with the {\tt cflux} command in XSPEC. 
We have to notice that, because {\it XMM-Newton} is unable to resolve M~81 X--1 from M~82 X--2, great caution is needed in order to interpret the results obtained from the spectral analysis.
We find very low values for the photon indices from the high-energy power-law (${\Gamma}=0.6-1.1$). These values are much lower than those typically seen in the high-energy spectra from ULXs (${\Gamma}=1.5-2.5$)
and might be a consequence of the fact that there is a large overlap between the Point Spread Functions (PSD) from M~82 X--1 and M~82 X--2 with {\it XMM-Newton}. Indeed, in studies of the {\it Chandra} spectra, it has been
found that the best description of the spectrum from M~82 X--2 is composed by multi-color disc plus power-law high-energy emission, the last with a very flat photon index, i.e. ${\Gamma}=1.19{\pm}0.06$.
This prevents us from making a quantitative comparison of the photon index/flux from the high-energy emission versus the frequency of the QPO, as in 
the relationships typically seen in stellar-mass BHBs \citep{vignarca03,titarchuk04}. The summed total (0.3-10\,keV) unabsorbed fluxes observed are in the range $(3.4-4.3){\times}10^{-11}\,{\rm erg}{\rm s}^{-1}{\rm cm}^{-2}$, consistently 
with what has been found
previously \citep{mucciarelli06}. In the case of Obs.~2,~5 we see an increase in the total flux, that is accompanied by a strong flattening of the spectra. This might indicate that in Obs.~2,~5 the 
contribution to the total flux from M~82 X--2 is important.  

\begin{table*}
 \centering
 \begin{minipage}{120mm}
  \caption{Results obtained from the spectral analysis.}
  \label{log_spe}
  \begin{tabular}{@{}lcccccc@{}}
  \hline
   Spectral parameter~$^1$      &   Obs.~1\,(Black)             &          Obs.~2\,(Red)        & Obs.~3\,(Green)                  &    Obs.~4\,(Dark Blue)        &    Obs.~5\,(Light Blue)              \\
 \hline
                                &                               &                               &      A                           &                               &                                      \\
 \hline
 ${\rm N}_{\rm H}$(1)\,$({\times}10^{22})\,({\rm cm}^{-2})$    &  $0.075{\pm}0.003$            &   $=$                         &   $=$                            &   $=$                         &   $=$                                \\
 ${\rm kT}_{1}$\,(keV)                                         &  $0.75{\pm}0.02$              &   $=$                         &   $=$                            &   $=$                         &   $=$                                \\
 ${\rm N}_{\rm H}$(2)\,$({\times}10^{22})\,({\rm cm}^{-2})$    &  $1.39{\pm}0.02$              &   $=$                         &   $=$                            &   $=$                         &   $=$                                \\
 ${\rm kT}_{2}$\,(keV)                                         &  $0.91{\pm}0.02$              &   $=$                         &   $=$                            &   $=$                         &   $=$                                \\
 ${\rm N}_{\rm H}$(3)\,$({\times}10^{22})\,({\rm cm}^{-2})$    &  $0.756{\pm}0.007$            &   $=$                         &   $=$                            &   $=$                         &   $=$                                \\
 ${\rm kT}_{\rm max}$\,(keV)                                   &  $0.18{\pm}0.17$              &  $0.29{\pm}0.05$              &  $0.19{\pm}0.02$                 &  $0.16{\pm}0.02$              &  $0.26{\pm}0.02$                     \\
  ${\rm N}_{\rm disc}$                                         & $70{\pm}23$                   &  $2.8{\pm}1.9$                &  $240_{-150}^{+230}$             &  $540{\pm}70$                 & $42{\pm}20$                          \\
  ${\rm E}_{\rm c}$\,(keV)                                     &  $5.38{\pm}0.10$              &   $4.49{\pm}0.10$             &   $6.75{\pm}0.12$                &   $6.22{\pm}0.12$             &  $6.04{\pm}0.10$                       \\
  ${\rm E}_{\rm f}$\,(keV)                                     &  $5.8{\pm}0.6$                &   $5.4{\pm}0.4$               &  $4.6{\pm}0.7$                   &   $4.5{\pm}0.8$               &  $5.6{\pm}0.7$                       \\
  ${\Gamma}$                                                   &  $0.86{\pm}0.02$              &   $0.60{\pm}0.02$             &  $1.13{\pm}0.02$                 &   $1.06{\pm}0.02$             &  $0.71{\pm}0.02$                     \\
  ${\chi}^{2}/{\nu}=849/755$                                   &                               &                               &                                  &                               &                                      \\
   ${\rm F}_{\rm pow}$\,$^2$\,(1-10\,keV)                      &  $1.27{\pm}0.06$              &  $1.58{\pm}0.08$              &   $0.93{\pm}0.05$                &  $1.17{\pm}0.06$              &  $1.89{\pm}0.10$  \\
                                &                               &                               &      B                           &                               &                                      \\
 \hline
 ${\rm N}_{\rm H}$(1)\,$({\times}10^{22})\,({\rm cm}^{-2})$    &  $0.09{\pm}0.02$              &   $=$                         &  $=$                             &   $=$                         &   $=$                                \\
 ${\rm kT}_{1}$\,(keV)                                         &  $0.76{\pm}0.02$              &   $=$                         &  $=$                             &   $=$                         &   $=$                                \\
 ${\rm N}_{\rm H}$(2)\,$({\times}10^{22})\,({\rm cm}^{-2})$    &  $1.47{\pm}0.02$              &   $=$                         &  $=$                             &   $=$                         &   $=$                                \\
 ${\rm kT}_{2}$\,(keV)                                         &  $0.92{\pm}0.02$              &   $=$                         &  $=$                             &   $=$                         &   $=$                                \\
 ${\rm N}_{\rm H}$(3)\,$({\times}10^{22})\,({\rm cm}^{-2})$    &  $0.84{\pm}0.02$              &   $=$                         &  $=$                             &   $=$                         &   $=$                                \\
 ${\rm kT}_{\rm max}$\,(keV)                                   &  $0.16{\pm}0.02$              &  $0.11{\pm}0.02$              &  $0.17{\pm}2$                    &  $0.15{\pm}0.02$              &  $0.23{\pm}0.02$                     \\
  ${\rm N}_{\rm disc}$                                         & $500^{+600}_{-300}$           &  $1\,600{\pm}1\,500$          &  $690{\pm}120$                   &  $1\,100{\pm}500$             & $100^{+60}_{-40}$                    \\
 ${\rm kT}_{\rm e}$\,(keV)                                     & ${\le}2$                      &  ${\le}2$                     &  $2.12{\pm}0.02$                 &  ${\le}2$                     & $2.18{\pm}0.06$                      \\
 ${\tau}$                                                      & $14.2{\pm}0.2$                &  $16.00{\pm}0.17$             &  $11.9{\pm}.5$                   &  $12.8{\pm}0.2$               & $15.1{\pm}0.7$                       \\
  ${\chi}^{2}/{\nu}=872/760$                                   &                               &                               &                                  &                               &                                      \\
 \hline
   ${\rm F}_{\rm compTT}$\,$^2$\,(1-10\,keV) &  $1.25{\pm}0.06$         &  $1.66{\pm}0.08$  &   $0.92{\pm}0.05$  &  $1.18{\pm}0.06$   &  $1.89{\pm}0.09$  \\
   ${\rm L}_{\rm X, compTT}$\,(0.3-10\,keV)  &  $2.45{\pm}0.13$         &  $3.21{\pm}0.16$  &   $1.83{\pm}0.98$  &  $2.35{\pm}0.11$   &  $3.60{\pm}0.18$  \\
\hline
\end{tabular}
\footnotetext{$^1$ The spectral models used are: A) {\tt TBabs*apec + TBabs*apec + TBabs(diskbb+highecut*powerlaw)} and B) {\tt TBabs*apec + TBabs*apec + TBabs(diskbb+compTT)}. Errors on parameters 
are $1{\sigma}$ errors and colors refer to the ones used in the energy spectra from Fig.~\ref{plot_spe}. }
\footnotetext{$^2$ Unabsorbed flux in units of $10^{-11}{\rm erg}\,{\rm s}^{-1}{\rm cm}^{-2}$. }
\footnotetext{Description of the parameters:\\
1) Column densities from each component of the diffuse emission from the galaxy ${\rm N}_{\rm H}$(1,2); 2) temperatures from the emission components describing the diffuse emission from the galaxy (${\rm kT}_{1}$ and ${\rm kT}_{2}$);
3) column density of the intrinsic emission from the source ${\rm N}_{\rm H}$(3); 4) temperature from the inner accretion disc (${\rm kT}_{\rm in}$) and normalization from the disc component (${\rm N}_{\rm disc}$); 5) power-law 
photon index (${\Gamma}$) and e-folding energy and cut-off (${\rm E}_{\rm f}$,${\rm E}_{\rm c}$) for the {\tt highecut*powerlaw} model component; 6) unabsorbed flux from the high-energy emission component 
(${\rm F}_{\rm pow}$ and ${\rm F}_{\rm compTT}$; for the {\tt highecut*powerlaw} or {\tt compTT} model components, respectively) in the 1-10\,keV energy range; 7) intrinsic luminosity of the high-energy component in 
the 0.3-10\,keV energy range and in units of ${\times}10^{40}{\rm erg}\,{\rm s}^{-1}$ (${\rm L}_{\rm compTT}$) assuming a distance of 3.9\,Mpc.
}
\end{minipage}
\end{table*}

\section{Results} \label{results}

In Fig.~\ref{plots_timing} we show the (1-10\,keV) PDS from the most recent set of 5 observations of the central region of M~82 performed with {\it XMM-Newton}. QPOs are seen in the PDS 
of Obs.~1--4 at frequencies ${\nu}=(48.1{\pm}1.1,8.3{\pm}0.2,44.1{\pm}0.5,35.2{\pm}1.2)$\,mHz, respectively. The best-fit parameters for the QPOs found are shown 
in Tab.~\ref{table_timing}. The detection significance of the QPOs of Obs.~1--3 is ${\ge}3.6{\sigma}$ 
(single trial) but the QPO of Obs.~4 is not significant. We have calculated the chance probability of getting these peaks by taking into
account the total number of trials and found them to be $8.7{\times}10^{-3},3.9{\times}10^{-2},4.2{\times}10^{-4}$ for Obs.~1,~2,~3, respectively. 
In order to estimate the number of trials for our detections we multiplied the number of observations analyzed, five, by the number of independent frequencies 
sampled. For the latter, we considered that we would have accepted QPOs in the range $0.01-1$\,Hz, and for each observation divided that range by the FWHM of the 
detected QPO. Therefore, after considering the number of trials, only the QPOs at Obs.~1,~3 are significant detections (${\approx}3{\sigma}$, after 
considering all the trials), whilst the QPO at Obs.~2 is detected only 
marginally (${\approx}2{\sigma}$). The frequency from the QPOs at Obs.1, 3 differ by ${\Delta}=4{\pm}1.2$\,mHz
and have a mean value of 46\,mHz. Although the difference between the two frequencies is statistically different from zero at the $3\,{\sigma}$ level. 
we will refer in the following to the QPOs from 
Obs.~1,~3 as the same one with a frequency equal to their mean value. The frequency of the QPO of Obs.~2 has much a lower frequency 
(${\nu}=8.3{\pm}0.2$\,mHz). 

\begin{figure}
\centering
 \includegraphics[bb=0 0 612 792,width=7cm,angle=270,clip]{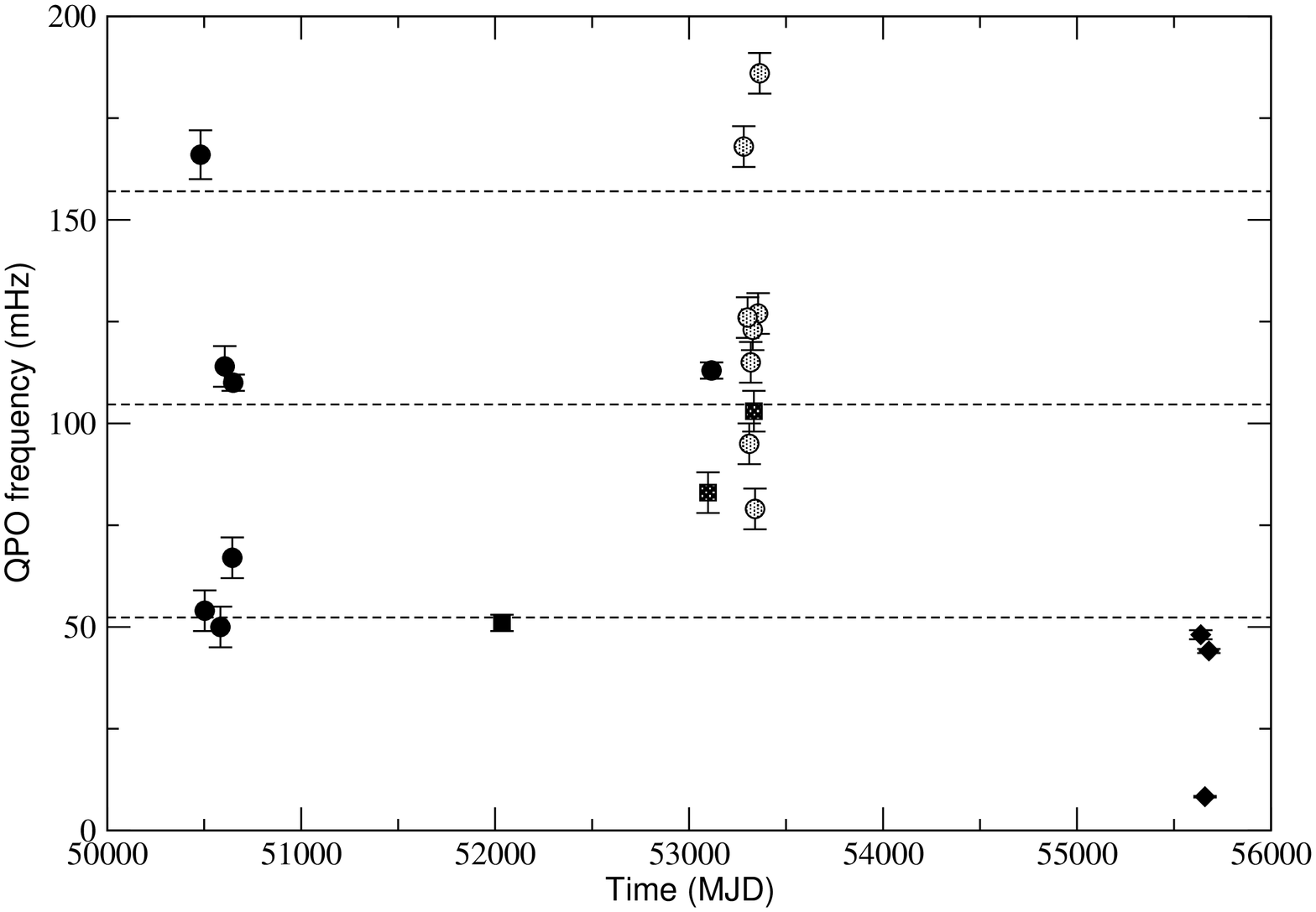}
 \caption{Time history of the centroid frequencies of the QPOs from M~82 with significance (${\sigma}>3$, single trial) from the total sample (see Tab.~\ref{log_qpos}). The QPOs from our work and 
from \citet{mucciarelli06,strohmayer03} are indicated with solid diamonds/circles/squares, respectively and the QPOs from \citet{kaaret06} are indicated with dotted-filled circles/squares for the 
non-significant/significant detections, after taking into account all the trials. Dashed lines mark the frequencies ${\nu}=(52.3,2{\times}52.3,3{\times}52.3)$\,mHz, in which ${\nu}=52.3$\,mHz
is the mean value of the frequencies from Tab.~\ref{log_qpos} in the range $(44-67)$\,mHz. }
 \label{qpos_time}
\end{figure}

We compared the QPOs found in our work with all those QPOs from M~82 X--1 that have been detected in the literature so far
(see list in Tab.~\ref{log_qpos}). The QPOs we have found at ${\nu}\approx46$\,mHz have roughly similar frequencies to the QPOs 
detected at ${\nu}{\approx}(50-70)$\,mHz by \citet{mucciarelli06}. On the other hand, the QPO we have detected at $(8.3{\pm}0.2)$\,mHz
has never seen before in M~82 X--1. In Fig.~\ref{qpos_time} we plot the time history of the centroid frequencies of all the QPOs 
detected from M~82 X--1 from all X-ray satellites (i.e. {\it XMM-Newton}, {\it Chandra} and {\it RXTE}) so far. 
QPOs marked with dot-filled circles were reported as ${\approx}4{\sigma}$ detections with {\it RXTE} by \citet{kaaret06}. 
The QPOs reported by \citet{kaaret06} appear to be single trial and no significances were given observation by observation. To include them in our comparison
we extracted the same observations and applied the same analysis as described in their paper, fitting 
the PDS and obtaining single-trial significances. We then derived final chance probabilities by estimating the number of trials as we had done for our detections.
We obtained that the chance probabilities of detection of their QPOs 
are in the range $0.268-0.873$ except for the QPOs at ($103{\pm}5,83{\pm}5$)\,mHz during MJD=53\,334,53\,098, for which we have found chance 
probabilities of $3.9{\times}10^{-2},6.31{\times}10^{-3}$. Therefore, applying the same significance criteria that we 
used for our observations, we only consider that the QPO at $(83{\pm}5)$\,mHz from \citet{kaaret06} is significant (${\approx}3{\sigma}$, after considering all the 
trials) whilst the QPO at $103{\pm}5$\,mHz is marginally significant (${\approx}2{\sigma}$) \footnote{In this work we consider as significant detections those with 
single-trial significance $>3{\sigma}$. In the most dubious cases (i.e. single-trial significance of $(3-4){\sigma}$) we also calculated the significance by 
considering all the trials.}. The second one is very similar to the QPO found previously from M~82 X--1 \citep{strohmayer03,fiorito04,mucciarelli06}. 
Nevertheless, a similar QPO to the first one, centred at ${\approx}70$\,mHz has been tentatively detected previously in a {\it Chandra} PDS of M~82 X--2 \citep{kaaret06}. Taking 
these considerations into account and looking only at the significant detections from M~82 X--1 (i.e. looking at the solid-filled circles from Fig.~\ref{qpos_time} only) we see a rough 
1:2:3 distribution in frequency of the QPOs, compatible with the QPO at ${\nu}{\approx}50$\,mHz being the 
fundamental one. In Fig.~\ref{qpos_time} we plot the distribution on time of the detections, from which the 1:2:3 distribution of the frequencies of the QPOs in M~82 X--1 is clearly seen.

\begin{figure}
\centering
 \includegraphics[bb=41 -18 564 700,width=6cm,angle=270,clip]{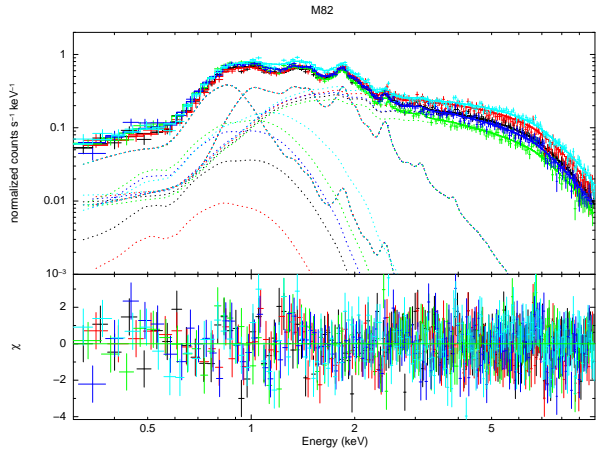}
 \caption{Energy spectra from M~82 during Obs 1--5 (black, red, green, dark and light blue, respectively).}
 \label{plot_spe}
\end{figure}

\begin{figure}
\centering
 \includegraphics[bb=0 0 612 792,width=7cm,angle=270,clip]{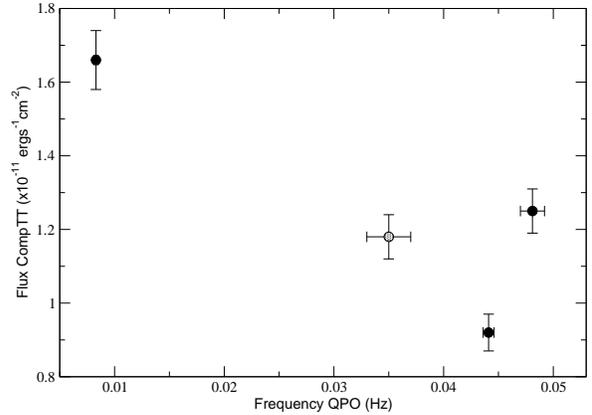}
 \caption{Unabsorbed flux from the high-energy component versus the frequency of the QPOs from our work (both in the 1-10\,keV energy range) detected with significance ${\ge}3.6{\sigma}$ (solid-filled circles). The QPO 
at ${\approx}30$\,mHz detected in our work with significance $<3{\sigma}$ has also been plotted (dot-filled circle).}
 \label{plot_anticorr}
\end{figure}

In Fig.~\ref{plot_spe} we plot all our spectra from {\it XMM-Newton}. Considering the most physical model (i.e. model B in Tab.~\ref{log_spe}) we see a change
in the flux from the high-energy component. The flux variations of the high-energy component from M~82 X--1 in Obs.1--4 
are $>3{\sigma}$ with respect to the mean value. We plotted the flux from the high-energy component versus the frequency of the QPO (see 
Fig.~\ref{plot_anticorr}). We can see that the appearance of a lower-frequency QPO coincides with an
increase of the flux from the high-energy component, i.e. a hardening of the spectrum. We discussed in Sec.~\ref{spectr} that this hardening is probably due to an increased flux 
from the nearby ULX M~82 X--2.  

\begin{table}
 \centering
 \begin{minipage}{120mm}
  \caption{List of all the QPOs from M~82 detected in the literature.}
  \label{log_qpos}
  \begin{tabular}{@{}lcccc@{}}
  \hline
   Date                      &  MJD     & QPO ${\nu}$\,(mHz)     &    ${\rm n}_{\sigma}$    &   Ref.~$^1$    \\
 \hline
 1997 Feb 02  &  50\,481  &    166$\pm$6          &  6.8   &  1  \\
 1997 Feb 24  &  50\,503  &     54$\pm$5          &  5.3   &  1  \\
 1997 May 16  &  50\,584  &     50$\pm$5          &  6.0   &  1  \\
 1997 Jun 07  &  50\,606  &    114$\pm$5          &  5.8   &  1  \\
 1997 Jul 16  &  50\,645  &     67$\pm$5          &  6.5   &  1  \\
 1997 Jul 21  &  50\,650  &    110$\pm$2          &  4.3   &  1  \\
 2004 Apr 21  &  53\,116  &    113$\pm$2          &  8.9   &  1  \\
 2001 May 06  &  52\,035  &     51$\pm$2          &  8.0   &  2  \\
 2004 Dec 25  &  53\,364  &   186$\pm$5           &  4.0   &  3  \\
 2004 Dec 17  &  53\,356  &   127$\pm$5           &  4.0   &  3  \\
 2004 Dec 01  &  53\,340  &    79$\pm$5           &  4.0   &  3  \\
 2004 Nov 25  &  53\,334  &   103$\pm$5           &  4.0   &  3  \\
 2004 Nov 19  &  53\,328  &   123$\pm$5           &  4.0   &  3  \\
 2004 Nov 09  &  53\,318  &   115$\pm$5           &  4.0   &  3  \\
 2004 Nov 01  &  53\,310  &    95$\pm$5           &  4.0   &  3  \\
 2004 Oct 24  &  53\,302  &   126$\pm$5           &  4.0   &  3  \\
 2004 Oct 04  &  53\,282  &   168$\pm$5           &  4.0   &  3  \\
 2004 Apr 03  &  53\,098  &    83$\pm$5           &  4.0   &  3  \\
 \hline
\end{tabular}
\footnotetext{$^1$ Bibliographic references:\\
(1) \citet{mucciarelli06};\\
(2) \citet{strohmayer03} and \\
(3) \citet{kaaret06}
}
\footnotetext{Notes: ${\rm n}_{\sigma}$ means the single trial detection significance \\
of the QPO. $1{\sigma}$ errors are used in the values from the \\
parameters.}
\end{minipage}
\end{table}

\section{Discussion}   \label{discuss}

In this work we report on the finding of two QPOs at a mean frequency value of ${\approx}46$\,mHz in the observations of the Ultra-Luminous X-ray source M~82 X--1. 
The {\it XMM-Newton}  observations have been performed in the time interval of ${\approx}250$\,days, over 
which we detected (at different significance levels) QPOs with different frequencies, i.e. a QPO at 46\,mHz in Obs.~1,~3, and a QPO at 8\,mHz during Obs.~2. These 
three observations have been performed in the time 
interval of only ${\approx}50$\,days, indicating the fast variability in this source can change on a scale of only a few weeks. For Obs.~4,~5 we do not detect any significant QPO 
(we obtained $3{\sigma}$ upper limits of $5\%$ for the fractional rms of a QPO at ${\approx}46$\,mHz for these observations). Additionally, 
we find that the QPO frequencies are roughly consistent with being anti-correlated with the flux from the high-energy component. Despite its low statistical 
significance, we refer to the last paragraph of this Section for the discussion of the 8 milli-hertz QPO and
the possibility that it is coming from the nearby ULX M~82 X--2. 

Comparing the QPO at ${\approx}46$\,mHz with previous findings (\citealt{mucciarelli06,strohmayer03,kaaret06}; see Tab.~\ref{log_qpos}) 
we see that all the detections have been distributed roughly in the 1:2:3 ratio (see Fig.~\ref{qpos_time}), with exception of the QPOs found 
by \citet{kaaret06}. \citet{kaaret06} found QPOs distributed rather uniformly over the frequency range ${\approx}(80-200)$\,mHz. This is in contrast
from what has been found by \citet{mucciarelli06}, where these QPOs are roughly equally spaced in frequency in a 1:2:3 ratio with the QPO at ${\approx}50$\,mHz 
as the fundamental one. As described in Sec.~\ref{results} we saw that, applying the same significance criteria that we used for our observations, only two QPOs 
from \citet{kaaret06} can be considered as significant (the second only marginal) detections, one at ${\approx}80$\,mHz and the other at ${\approx}100$\,mHz (see Fig.~\ref{qpos_time}). A 
QPO at a frequency similar to ${\approx}80$\,mHz has been tentatively identified in a {\it Chandra} observation
as coming from M~82 X--2. The QPO at ${\approx}100$\,mHz is similar to what has been found previously in the case of M~82 X--1. Looking at Fig.~\ref{qpos_time}
we can see that the frequencies of the QPOs cluster at around the frequencies ${\nu}=(52.3,2{\times}52.3,3{\times}52.3)$\,mHz, in which ${\nu}=52.3$\,mHz
is the mean value of the frequencies from Tab.~\ref{log_qpos} in the range $(44-67)$\,mHz, that we will refer to as the fundamental QPO frequency hereafter. The 
lines In Fig.~\ref{qpos_time} (which would correspond to the fundamental QPO, and twice and three times this value)
show that the values for the first and second group of frequencies are closer to the expected harmonic values than the spread of the points around the fundamental. Nevertheless, it 
is difficult to assess for the significance of the clustering of the QPOs around the previous frequencies, due to the scarcity of points.  

Previously, \citet{fiorito04} pointed out
to the 1:2 frequency ratio for the QPOs at frequencies ${\approx}50,100$\,mHz, using all the available data at that time. We suggest with our analysis (together with 
the results from the previous analysis by \citealt{strohmayer03} and \citealt{mucciarelli06}) that the QPOs from M~82 X--1 are in fact roughly distributed harmonically in 
the 1:2:3 ratio, taking all the available data so far. The final proof would be to detect all these harmonic QPO simultaneously but, as discussed in the next paragraph, that coincidence is unlikely, given the detection 
limits from the current instrumentation and/or length of the current observations. 

We have found that the frequency of the QPOs from M~82 X--1 do not appear to be distributed randomly, but seem to prefer special values, although
the number of detections is too low to prove it. In addition to this, 
these values are also compatible with being simple multiples of each other. Both facts lead to the suggestion that
the QPOs in M~82 X--1 could be interpreted as the first identification of HFQPOs in a ULX, in analogy to 
the HFQPOs already identified in BHBs. Nevertheless, a major difference with the case of BHBs is that in the latter the fundamental frequency has never
been observed. The characteristics of our QPOs (high values for the coherence and fractional rms) and their location at the high-frequency break of the low-frequency noise 
have some similarities with the type-C LFQPOs found in BHBs \citep{casella05}. Nevertheless, these characteristics do not exclude other classifications, since we 
are only able to detect the brightest signals. HFQPOs in BHBs are signals at frequencies 35--450\,Hz observed in a few systems. These correspond to the time scales expected from 
the Keplerian motion in the innermost regions of the gravitational well of the BH. There is evidence that in BHBs these oscillations appear at specific frequencies and, in 
a few cases, pairs of peaks have been observed simultaneously. These pairs of peaks have been interpreted associating the QPO frequencies to relativistic time scales at a specific 
radius, where these frequencies are in resonance, resulting therefore in special frequency ratios \citep{kluzniak01}. Nevertheless, other models (see \citealt{stella99}
and references therein) make predictions about frequencies of the high-frequency peaks, but do not lead to specific ratios. In the case of M~82 X--1, as \citet{fiorito04} pointed 
out, it could be that the presence of one dominant QPO, i.e. 50 or 100 or 150\,mHz in different observations, is not an unusual occurrence and can be explained 
as the result of the local driving frequency conditions in the coronal region. This leads to the fact that a resonance condition is established for one particular eigenmode of the 
compact coronal region so that this mode is predominantly observed.

If the QPO detected in M~82 X--1 is a fundamental HFQPO, then it appears at much lower frequency (${\approx}50$\,mHz) than those observed in BHBs (35--450\,Hz), by three orders of 
magnitude. Scaling linearly with the mass of a stellar-mass BH ($10\,{\rm M}_{\odot}$) if these QPOs have the same origin, the frequencies we have found lead to a mass
of ${\rm M}_{\rm BH}{\approx}(10^4-10^{5})\,{\rm M}_{\odot}$ (taking into account the whole range of possible values of the spin of the BH) for the mass of the BH in M~82 X--1. Other 
mass estimates, this time based on spectral-timing scaling relationships from systems of different mass \citep{titarchuk04}, have provided a mass estimate 
of ${\approx}10^3\,{\rm M}_{\odot}$ \citep{fiorito04}. Nevertheless, caution is required in order to apply relationships based solely on the mass of the BH systems. As
already pointed out by several authors (\citealt{mchardy06} for AGNs; \citealt{soria07} and \citealt{casella08} for ULXs) the accretion rate is an important parameter that, together with
the mass of the BH, should be considered as the main drivers in these scaling relationships. This effect could lead to a much smaller mass for the
BH, i.e. \citet{soria07} and \citet{casella08}. In addition, it remains to date unclear how timing properties and BH masses are related, and what are the properties of the
accretion states in ULXs compared to those of Galactic BHBs. 

Regarding the low-frequency (${\approx}8$\,mHz) QPO, it is unlikely that it is an analogous to the HFQPOs observed in BHBs, since its frequency is much lower 
than ${\approx}50$\,mHz. If it was the fundamental harmonic analog to the HFQPOs observed in BHBs, many harmonics should have been observed, but none has been detected so far.
The low frequency of this QPO would instead indicate an analogy to the Low-Frequency QPOs (LFQPOs) observed in BHBs. LFQPOs with frequencies
ranging from a few mHz to ${\approx}10$\,Hz are a common feature in almost all BHBs. Despite LFQPOs being known for several decades, their
origin is still not understood and there is no consensus about their physical nature. Nevertheless, any model describing the origin of LFQPOs associates them to the
inner parts of the flow around BHs. If this association is correct, scaling the frequency linearly with the possible values for the mass of M~81 X--1 ($10^2-10^4\,{\rm M}_{\odot}$) leads
to a frequency of 0.08-8\,Hz for the LFQPOs in stellar-mass ($10\,{\rm M}_{\odot}$) BH. This range of frequencies is compatible with those observed for LFQPOs in BHBs.
Milli-hertz QPOs in the range of (2-7)\,mHz have been previously detected in M~82 X--2 \citep{feng10}, with a coherence 
value (defined as the frequency of the QPO divided by its FWHM) of ${\approx}1.0$. The low-frequency QPO in our case is much narrower, with a coherence value 
of ${\approx}9$. Nevertheless, such high values of the coherence (i.e. narrow QPOs) are typical of the most common QPOs observed in almost all sources. In our case, as in
\citet{feng10} the detection of the milli-hertz occurs only when M~82 X--2 is luminous, above ${\approx}10^{40}\,{\rm erg}{\rm s}^{-1}$. This suggests the idea that the
milli-hertz QPO is from M~82 X--2. In order to test this further, in a way similar to \citet{feng07}, we extracted the flux from two separated halves of a circular region centred in M~82 X--1
for all the observations. We found that the QPOs from M~82 X--1 fall below the detection threshold in all the separate halves. But in the case of Obs.~2 we detect a QPO at ${\approx}8$\,mHz
only in the half circle which contains M~82 X--2. This strongly suggests that for this observation the observed signal is associated to M~82 X--2.

\section*{Acknowledgments}

We thank the anonymous referee for helpful comments. This work is based on observations made with XMM-Newton, an ESA science mission with instruments and contributions directly
funded by ESA member states and the USA (NASA). MCG acknowledges support from INAF through a 2010 postdoctoral fellowship. 
TB acknowledges support from grant ASI-INAF I/009/10/. LZ acknowledges financial support from INAF through grant PRIN-2011-1
("Challenging Ultraluminous X-ray sources: chasing their black holes and formation pathways") and ASI/INAF grant no. I/009/10/0.
The research leading to these results has received from the European Community’s Seventh Framework Programme (FP7/2007-2013) under
grant agreement number ITN 215212 Black Hole Universe.

\end{document}